\documentstyle[aps,tighten]{revtex}

\begin{document}
\draft

%\twocolumn[\hsize\textwidth\columnwidth\hsize\csname@twocolumnfalse%
%\endcsname

\title {Separable-entangled frontier in a bipartite harmonic system}

\author{Constantino Tsallis $^{(a)}$\thanks{tsallis@cbpf.br, 
prato@mail.famaf.unc.edu.ar, celia@cbpf.br}, 
Domingo Prato $^{(a,b)}$ and Celia Anteneodo $^{(a)}$}

\address{$^{(a)}$Centro Brasileiro de Pesquisas Fisicas, 
Xavier Sigaud 150, 22290-180, 
Rio de Janeiro-RJ, Brazil \\
$^{(b)}$FaMAF, Universidad Nacional de Cordoba, Ciudad Universitaria, 
5000 Cordoba, Argentina\\
}

\maketitle

\begin{abstract}

We consider a statistical mixture of two identical harmonic oscillators which is characterized by four 
parameters, namely, the concentrations ($x$ and $y$) of diagonal and nondiagonal 
bipartite states, and their associated thermal-like noises ($T/\alpha$ and $T$, 
respectively). The fully random mixture 
of two spins 1/2 as well as the Einstein-Podolsky-Rosen (EPR) state are recovered as particular instances. 
By using the conditional nonextensive entropy as introduced by Abe and Rajagopal, we calculate the 
separable-entangled frontier. Although this procedure is known to provide a necessary but in general {\it not} 
sufficient condition for separability, it does recover, in the particular case $x=T=0$ ($\forall \alpha$), 
the 1/3 exact result known as  Peres' criterion. This is an indication of reliability of the calculation 
of the frontier in the entire parameter space. The $x=0$ frontier remarkably resembles to the critical 
line associated with standard diluted ferromagnetism where the entangled region corresponds to the ordered 
one and the separable region to the paramagnetic one. The entangled region generically shrinks for increasing 
$T$ or increasing $\alpha$. 

\end{abstract}


%\begin{multicols}{2}

%\bigskip

Quantum entanglement is a quite amazing physical phenomenon, and has attracted intensive interest in recent 
years due to its possible applications in quantum computation, teleportation and cryptography, as well as to 
its connections to quantum chaos \cite{werner,GHZ,ekert,bennett,zurek2,quantumchaos,popescu,barenco,zurek,horodecki1,horodecki2,peres,horod,horodeckicube,popescu2,horodecki3,horodecki4,lloyd,bruss}. A nonextensive statistical mechanics \cite{tsallis} was proposed in 
1988 by one of us, and is currently applied \cite{levy,logistic,rafelski,beck1,ion,beck2} to a variety 
of thermodynamically anomalous systems which, in one way or another, exhibit (multi)fractals aspects. Among 
these anomalous systems, a prominent place is occupied by systems including long-range interactions and L\'evy 
distributions. So being, it is after all not surprising that this thermostatistical formalism has interesting 
implications \cite{aberajagopal,others,sethmichel} in the area of quantum entanglement and its intrinsic 
nonlocality, thus showing the  confluence of two concepts coming from distinct physical areas.

Quantum systems can be more or less entangled, which makes relevant the discussion of whether a 
given system is or not separable. Separability, which we shall define in detail later on, is a crucial 
feature in the discussion on whether a quantum physical system is susceptible of a {\it local realistic} 
description with hidden variables. These issues were first discussed in 1935 by Einstein, Podolsky and 
Rosen (EPR) \cite{EPR} and by Schroedinger \cite{schroedinger}, and since then by many others 
\cite{werner,GHZ,ekert,bennett,zurek2,quantumchaos,popescu,barenco,zurek,horodecki1,horodecki2,peres,horod,horodeckicube,popescu2,horodecki3,horodecki4,lloyd,bruss}.
 As we have mentioned above we shall see that entropic nonextensivity provides a path through which it 
is possible to discuss quantum entanglement \cite{aberajagopal,sethmichel,walter}.

In this work we analyze a composite system of two harmonic oscillators with identical energy spectra. Following 
along the lines of Abe and Rajagopal \cite{aberajagopal}, we use the nonextensive entropy $S_q \equiv 
\frac{1- Tr \rho^q}{q-1}\;(q \in \cal{R}; \; $$S_1$$= -$ $Tr$ $ \rho \ln \rho)$, $\rho$ being the density
 matrix, in order to study the frontier between {\it separable} ({\it quantum nonentangled}) and 
{\it nonseparable} ({\it quantum entangled}) regions. More precisely, we determine a frontier in some 
parameter-space which is either the exact one or an overestimation of the separable region. We remind that the 
entropic arguments that are used within this approach provide, like the Peres' partial transpose method 
\cite{peres}, necessary but {\it not} sufficient conditions for separability. The mixed state we shall 
consider for this bipartite system involves Boltzmann-like probabilities in such a way that the bipartite 
spin 1/2 system is recovered as a particular case. More precisely, the 1/3 Peres' criterion for separability 
will emerge as the $T=0$ limit of this specific mixed state. An increase of the temperature $T$ incorporates 
higher energy levels in the mixed state under consideration, which has the effect of enlarging the separable region. 

In order to make this work self-contained we will present, in the following, the connection between the 
nonextensive statistical mechanics and quantum entanglement. 
Let us begin with the definition of a generalized entropy for a quantum system

\begin{equation}
\label{Sq}
S_q=\frac{1- Tr \rho^q}{q-1}\;\;\;(q \in \Re; \; Tr \rho=1;\;S_1=-Tr\;\rho \ln \rho)\;, 
\end{equation}
where $\rho$ is the density operator.
Let us assume now that $\rho$ is the density operator associated with a composed system $A+B$. Then the marginal 
density operators are given by  $\rho_A \equiv Tr_B\; \rho$ and $\rho_B \equiv Tr_A\; \rho$ ($Tr_A \;\rho_A=
Tr_B \;\rho_B=1$). The systems $A$ and $B$ are said to be {\it uncorrelated } (or {\it independent}) if and only if 

\begin{equation}
\rho = \rho_A \otimes \rho_B \;.
\end{equation}
Otherwise they are said to be {\it correlated}. Two correlated systems are said to be {\it separable} (or {\it 
unentangled}) if and only if it is possible to write $\rho$ as follows:
 
\begin{equation}
\rho = \sum_{i=1}^W p_i \;\rho_A^{(i)} \otimes \rho_B^{(i)} \;\;\;\; (p_i \ge 0\; \forall i; \;\sum_{i=1}^W p_i=1).
\end{equation}
The limiting case of independency is recovered for {\it certainty}, i.e., if all $\{p_i\}$ vanish excepting one which 
equals unity. {\it Nonseparability} (or {\it entanglement}) is at the basis of the amazing phenomena mentioned before
 and, as already pointed, at the center of the admissibility of a {\it local realistic} 
description of the system in terms of hidden variables. 
As it is known correlation is a concept which is meaningful both classically and quantically. Entanglement, 
in the present sense, is meaningful only within quantum mechanics. The characterization of quantum entanglement is 
not necessarily simple to implement, since it might be relatively easy in a specific case to exhibit the form of 
Eq. (3), but it can be nontrivial to prove that it {\it cannot} be presented in that form. Consequently, along the 
years appreciable effort has been dedicated to the establishment of general operational criteria, preferentially 
in the form of necessary and sufficient conditions whenever possible. Peres\cite{peres} pointed out a few years ago a 
{\it necessary} condition for separability, namely the nonnegativity of the partial transpose of the density matrix. 
In some simple situations (like the simple mixed state of two spin 1/2) Peres' criterion 
is now known to also be a {\it sufficient} condition. But, as soon as the case is slightly more complex (e.g., 
$3 \times 3$ or $2 \times 4$ matrices) it is known now to be {\it insufficient}.
Within this scenario Abe and Rajagopal\cite{aberajagopal} recently proposed a different condition, 
claimed to be a {\it necessary} one, based on the nonextensive entropic form $S_q$ given in Eq. (1). Let us summarize 
the idea. The quantum version of conditional probabilities is not easy to formulate in spite of being so simple within 
a classical framework. The difficulties come from the fact that generically $\rho$ does not commute 
with either $\rho_A$ or $\rho_B$. Consistently, the concept of {\it quantum conditional entropy} is a sloppy one. 
Abe and Rajagopal suggested a manner of shortcutting this difficulty, namely through the adoption, of the following 
definitions of the conditional entropies $S_q(A|B)$ and $S_q(B|A)$, respectively given by

\begin{equation} \label{SqAB}
S_q(A|B)\equiv \frac{S_q(A+B)-S_q(B)}{1+(1-q)S_q(B)},
\end{equation}
and
\begin{equation} \label{SqBA}
S_q(B|A)\equiv \frac{S_q(A+B)-S_q(A)}{1+(1-q)S_q(A)},
\end{equation}
where $S_q(A+B) \equiv S_q(\rho)$, $S_q(A) \equiv S_q(\rho_A)$ and $S_q(B) \equiv S_q(\rho_B)$. Obviously, for the 
case of independence, i.e., when $\rho = \rho_A \otimes \rho_B$, these expressions lead to $S_q(A|B)=S_q(A)$ and 
$S_q(B|A)=S_q(B)$, known to be true also in quantum mechanics.

Both classical and quantum entropies $S_q(A+B)$, $S_q(A)$ and $S_q(B)$ are always nonnegative. This is not the case 
of the conditional entropies $S_q(A|B)$ and $S_q(B|A)$, which are always nonnegative classically, but which can be 
negative quantically (see also \cite{cerfadami}). It is therefore natural to expect that separability implies 
nonnegativity of the conditional entropies for all $q$. This is the criterion proposed in \cite{aberajagopal}. 

Peres' criterion seems to be less restrictive than the entropic one in the sense that it might provide a larger quantum entangled region. 
Although we have no proof that it cannot be the other way around, 
we have not encountered such an example.
In several cases, including some of increasingly many spins, both criteria have produced the same result, presumably the 
asymptotically exact answer. Details can be found in \cite{GHZ,horodecki2,peres,horod,bruss,alcaraz}. As a general trend, it seems 
that the transposed matrix criterion {\it overestimates the separable region} (i.e., {\it underestimates the 
quantum entangled region}) less or equally than the conditional entropy criterion does. 
It could well be that whenever both criteria produce the same result, this result is the exact answer \cite{horodeckitsallis}.

Equations (\ref{SqAB}) and (\ref{SqBA}) have in fact been proved for a classical system \cite{walter}, not for a quantum one. 
It seems however plausible \cite{aberajagopal} that they preserve the same form in both classical and quantum cases.
Such conditional entropies enable what is a necessary condition for separability.  
Imposing the conditions  $S_q(A|B), S_q(B|A)>0$ for all values of $q$, the  particular space of parameters becomes 
divided into two regions: one where any conditional entropy $S_q$ is positive and the other one
- a domain of entangled states - where at least one conditional entropy 
is negative for some $q$. Therefore, a line (or a frontier, generally speaking) like a critical one, emerges. 

Within this standpoint, it becomes 
clear that the Boltzmann-Gibbs-Shannon entropy ($q=1$) 
is a concept too poor for properly discussing quantum entanglement, a conclusion recently reached also by 
Brukner and Zeilinger\cite{others} from a different path. 
An appreciable amount of arguments are now available in the literature which connect quantum entanglement and 
thermodynamics \cite{zurek2,horodecki1,horodecki2,popescu2,horodecki3,horodecki4,others}. 

In what follows we consider, for the two oscillators $A$ and $B$, the basis 
$|n\rangle_A|m\rangle_B \equiv |n,m\rangle$, where $n,m=0,1,2,...$. Let us define the symmetric and antisymmetric 
states $|n,m^{\pm} \rangle \equiv \frac{1}{\sqrt 2}(|n,m\rangle \pm |m,n\rangle)$ for $n>m$. 
These states will play a role similar to that of the EPR state. The mixed state we are interested in is the following one:

\begin{eqnarray}
\rho_{A+B} = \frac{1-x-y}{4} \Bigl[|0,0\rangle\langle0,0| + |1,1\rangle\langle1,1| + 
|1,0\rangle\langle1,0| + |0,1\rangle\langle0,1|\Bigr]  \nonumber \\ 
+ x\;a \sum_{n=2}^{\infty} e^{- 2n\alpha/T} |n,n\rangle\langle n,n| 
+ y\;b \sum_{n=1}^{\infty} \sum_{m=0}^{n-1} e^{-(n+m)/T} |n,m^- \rangle\langle n,m^- | \;,
\label{roAB}
\end{eqnarray}
where $T \ge 0$, $\alpha \ge 0$,
\begin{equation}
a \equiv 1/{\displaystyle \sum_{n=2}^{\infty}} e^{-2n\alpha/T}=
(1- e^{-2 \alpha/T})\; e^{4 \alpha/T} 
\end{equation}
and
\begin{equation}
b\equiv 1/{\displaystyle \sum_{n=1}^{\infty} \sum_{m=0}^{n-1} } e^{-(n+m)/T}
=2             (1-e^{-1/T}) \sinh (1/T)  \;.
\end{equation}
We easily verify that $Tr \rho_{A+B}=1$. Since the eigenvalues of $\rho_{A+B}$ must be numbers within $[0,1]$, 
we have that $0 \le x,y \le 1$ and $x+y \le 1$. Parameter $\alpha$ measures the importance of the noise 
associated with the nondiagonal terms with respect to the noise associated with the diagonal ones. 
Parameter $T$ is a temperature-like measure of noise. Notice that, for $x=y=0$, we have 

\begin{equation}
\rho_{A+B} = \frac{1}{4} \Bigl[|0,0\rangle\langle0,0| + |1,1\rangle\langle1,1| 
+ |1,0\rangle\langle1,0| + |0,1\rangle\langle0,1|\Bigr]=\frac{1}{4}{\hat 1}_4\;,
\end{equation}
where ${\hat 1}_4$ is the 4-dimensional identity matrix. In other words, this state is isomorphic to a fully 
random state of two spins 1/2. Notice also that, for $T=0$ and $(x,y)=(0,1)$, we have $\rho_{A+B}= 
|1,0^-\rangle\langle1,0^-| = \frac{1}{\sqrt 2} \Bigl[|1,0\rangle\langle1,0| - |0,1\rangle\langle0,1| \Bigr]$, 
which is isomorphic to the celebrated EPR state. 
The separable-entangled frontier for the $T=x=0$ particular case is known
 \cite{peres,aberajagopal,sethmichel,walter}, namely $y=1/3\;(\forall \alpha)$.
Before we address the discussion of the full frontier in the $(x,y,T,\alpha)$-space, it is convenient to 
rewrite Eq. (\ref{roAB}) in the following equivalent form:

\begin{eqnarray}
\rho_{A+B} &=& \frac{1-x-y}{4} \Bigl[ |0,0\rangle\langle0,0| + |1,1\rangle\langle1,1| 
+ |1,0^+\rangle\langle1,0^+| \Bigr]  \nonumber \\
&+& x\;a \sum_{n=2}^{\infty} e^{-2n\alpha/T} |n,n\rangle\langle n,n|   
+ \Bigl(\frac{1-x-y}{4} +y\;b\; e^{-1/T}\Bigr) |1,0^-\rangle\langle1,0^-|    \nonumber \\ 
&+& y\;b \sum_{n=2}^{\infty} \sum_{m=0}^{n-1} e^{-(n+m)/T} |n,m^- \rangle\langle n,m^- | \;,
\end{eqnarray}
Due to the orthonormality of all the states appearing in this expression, the calculation of $(\rho_{A+B})^q$ 
is easily carried out. It follows

\begin{eqnarray}
Tr (\rho_{A+B})^q  
&=& 3\Bigl( \frac{1-x-y}{4}\Bigr)^q 
+ \frac{ x^q\,(1- e^{-2 \alpha/T})^q }{1-e^{-2q \alpha/T}}  \nonumber \\
&+&\Bigl(  \frac{1-x-y}{4} + 2y\,(1-e^{-1/T})\, e^{-1/T} \sinh (1/T) \Bigr)^q  \nonumber \\
&+&\frac{ \Bigl( 2y\,(1-e^{-1/T}) \,e^{-1/T} \sinh (1/T)\Bigr)^q \, (1+ e^{-q/T}- e^{-2q/T})}
{2\sinh(q/T) \,(1-e^{-q/T})}           
\end{eqnarray}

Let us now address the calculation of $(\rho_A)^q$. We have

\begin{eqnarray}
\rho_B \equiv Tr_A \rho_{A+B}
&=& \Bigl(\frac{1-x-y}{2} + y\,e^{-1/T} \sinh (1/T)\Bigr) |0\rangle\langle0|  \nonumber \\
&+& \Bigl(\frac{1-x-y}{2} + y\,\sinh(1/T)\,(e^{-1/T}-e^{-2/T}+e^{-3/T})\Bigr) |1\rangle\langle1|  \nonumber \\
&+& \sum_{n=2}^{\infty}  \Bigl( x\, (1-e^{-2 \alpha/T})\, e^{-2(n-2) \alpha/T}
+ y\,\sinh (1/T)\,(e^{-n/T}-e^{-2 n/T}+e^{-(2n+1)/T} ) \Bigr)
 |n\rangle\langle n| \;,
\end{eqnarray}
where $|n\rangle$ denotes the $n^{th}$ state of oscillator $B$. It now follows
\begin{eqnarray}
Tr(\rho_B)^q &=& \Bigl( \frac{1-x-y}{2} +
 y\, e^{-1/T} \sinh (1/T) \Bigr)^q \nonumber \\
&+&  \Bigl( \frac{1-x-y}{2} + y\, \sinh (1/T)\, 
(e^{-1/T} -e^{-2/T} + e^{-3/T})  \Bigr)^q   \nonumber \\
&+& \displaystyle \sum_{n=2}^{\infty}  
\Bigl( x\, (1-e^{-2 \alpha/T})\, e^{-2(n-2) \alpha/T} +y\,\sinh (1/T) \,
( e^{-n/T} - e^{-2n/T} + e^{-(2n+1)/T})  \Bigr)^q  
\end{eqnarray}
Now that we have $Tr(\rho_{A+B})^q $ and $Tr(\rho_B)^q $ as explicit 
functions of $(x,y,T,\alpha)$, we can immediately obtain  
$S_q(A|B) (=S_q(B|A))$ through the use of Eqs. (\ref{Sq}) and (\ref{SqAB}) yielding

\begin{equation}
S_q(A/B)=\frac{1}{q-1}\left( 1-\frac{Tr(\rho_{A+B})^q}{Tr(\rho_B)^q }\right) \, .
\end{equation}
$S_q(A|B)$ as a function of $(x,y)$ for typical values of $(q,T)$ and $\alpha=1$ is 
represented in Figs. 1 and 2. 
By comparing these figures, notice the nonuniform convergence at $T=1/q=0$.

Fig. 3 exhibits, in $(y,q)$ space, the line verifying $S_q(A|B)=0$, for particular values of 
$(x,T,\alpha)$. The physical region is the one satisfying $0 \le x+y \le 1$. 
Given $x$,  for $y$ above this critical line and within the physical region $y\leq 1-x$, 
one has $S_q(A|B)<0$, which signals the existence of an entangled state. 
Notice that for sufficiently small $x$, the strongest restriction to the separable region is imposed at $q\to\infty$. 
As $x$ increases, the minimum of $y$ occurs at a finite $q$, $q_{min}$ (the possible nonmonotonicity 
of $S_q(A|B)$ with regard to $q$ has already been noticed by \cite{wolf}). 
Since the strongest restriction to separability corresponds to the nonnegativity of both 
$S_{q_{min}}(A|B)$ and $S_{q_{min}}(B|A)$, then this condition is the one which better
approaches the separable-entangled frontier we are looking for and which is 
represented in Fig. 4.  
The entangled region in the $(x,y,T)$-space is illustrated in Fig. 5 for 
various values of $(x,\alpha)$.  
For $T=0$ and arbitrary 
$\alpha$, the system is entangled if $(1-x)/3 <y \le 1-x$, 
which immediately recovers the Peres' 1/3 criterion for $x=0$. 
As $T$ increases, the entangled domain shrinks against the border $x+y=1$ and 
dissapears at a critical $T$, $T_c$ ($T_c=1/\ln 2$ for $x=0$).

If the conditional entropy $S_q(A|B)$ is a monotonically decreasing function of $q$, then  
the strongest case corresponds to the nonnegativity of $S_{\infty}(A|B)$. 
Whenever the frontier is fully defined by $q\to\infty$, the system is entangled in the 
region defined by

\begin{equation}
y^*<y\le 1-x  \;\;\;\;\mbox{and}\;\;\;\; x<x^*  
\end{equation}
where
\begin{equation}
y^*  \equiv \frac{1-x}{3-2 \;e^{-1/T} (2+e^{-1/T} -2\; e^{-2/T})} 
\end{equation}
and 
\begin{equation}
x^*  \equiv \frac{\left( 2(1-e^{-2/T})(2-3 \;e^{-1/T} + e^{-3/T} - e^{-4/T})-1 \right) y +1 }{(5-4\; e^{-2\alpha/T})}
\end{equation}
However, for sufficiently small $T$, there is a curved line 
(corresponding to the nonlinear portion of the frontier in Fig. 4), 
given by $y=f(x)$, 
 which is defined from the conditional entropy for finite $q_{min}$ and that 
can be numerically determined. In this case, the entangled region is given by

\begin{equation}
y^*<y\le 1-x \;\;\;\;\mbox{and}\;\;\;\; y>f(x).
\end{equation}

An increase of the noise parameter $T$ (see Figs. 4 and 5), i.e.,  
an increase of the relative weight of the high energy states of the oscillators, 
causes shrinking of the 
entangled region. This is intuitively expected. Also, interestingly enough, decreasing the parameter $\alpha$ 
(see Figs. 4 and 5), 
which corresponds to an increase of the noise associated with the diagonal states, 
causes the entangled region to enlarge. This is not surprising after all, since this noise makes the diagonal terms 
to become less effective than the off-diagonal ones, which makes quantum coherence to be relatively stronger. 
The interplay of these effects may cause an interesting re-entrance, as illustrated in Fig. 5(a) for the value $x=0.6$. 
In this situation, corresponding in fact to small values of $\alpha$, it is possible to loose the quantum entanglement 
as $T$ increases, and then recover it back, until definitive loss for very large $T$.
To conclude, we believe that the present results provide quantitative insight on the influence of the external 
world (here represented by two different types of noise) on a quantum device. 

One of us (DP) acknowledges warm hospitality at the Centro Brasileiro de Pesquisas F\'{\i}sicas. 
Also, we acknowledge partial financial support from PRONEX, CNPq and FAPERJ.

\vspace*{1cm}

\centerline{{\bf FIGURE CAPTIONS}}

\bigskip
\noindent
Fig. 1.~
$S_q(A|B)$ as a function of $(x,y)$ for $\alpha=1$, $T=0.1$ and typical values of $q$.

\bigskip
\noindent
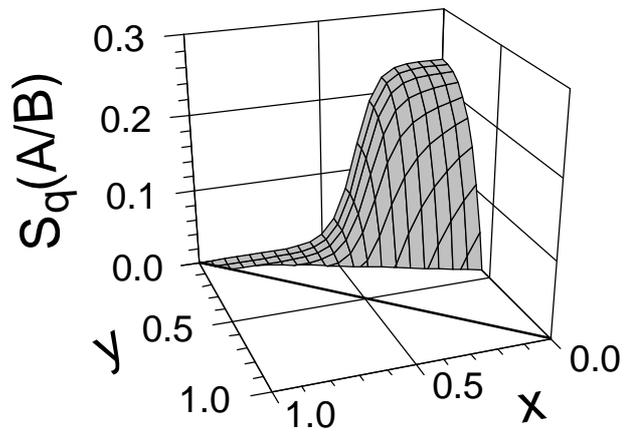
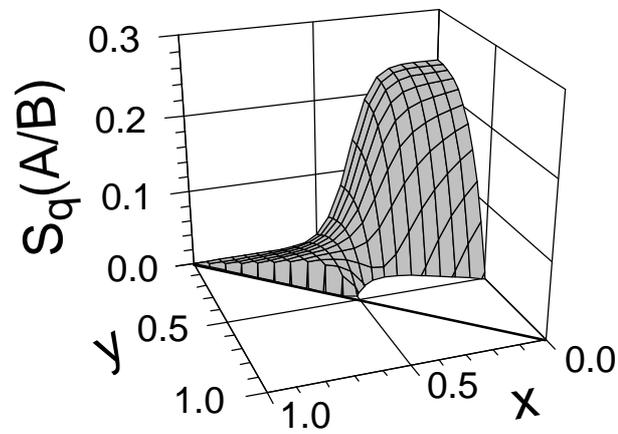
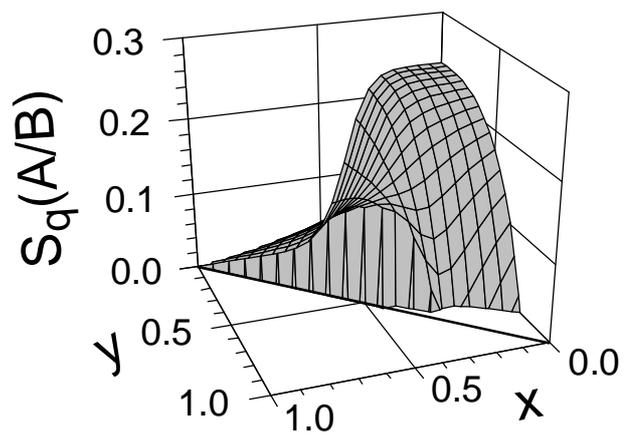
Fig. 2.~
$S_q(A|B)$ as a function of $(x,y)$ for $\alpha=1$, typical values of $T$ and $q=5$.

\bigskip
\noindent
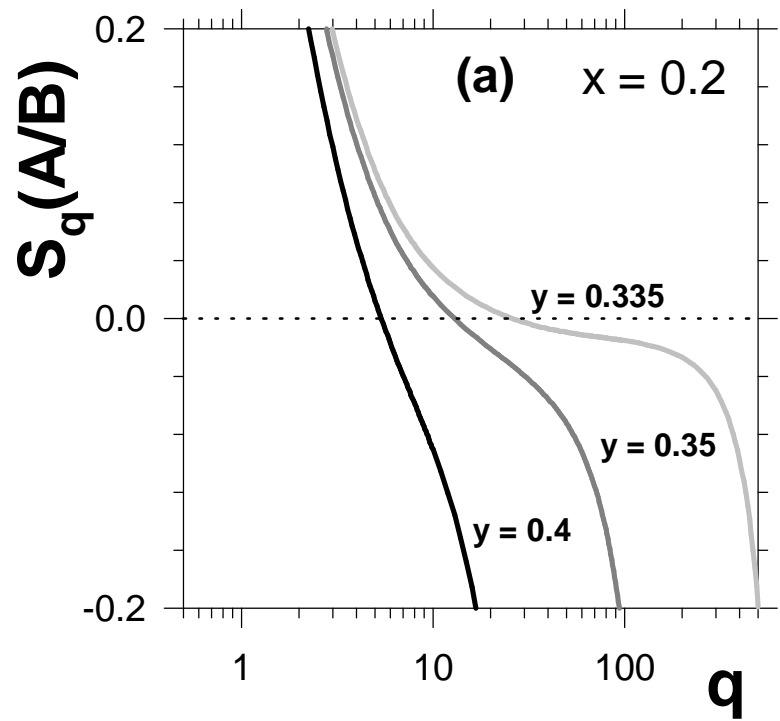
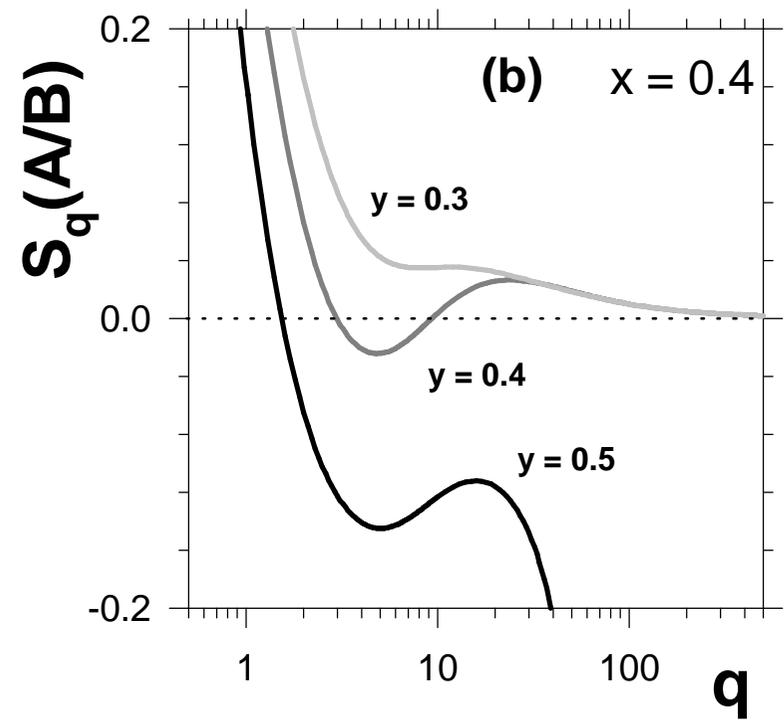
Fig. 3.~
The line $S_q(A|B)=0$, in $(y,q)$-space,  for particular values of $x$, $T=0.5$ 
and $\alpha=1$.

\bigskip
\noindent
Fig. 4.~
Frontier in the $(x,y)$-space for typical values of $T$ and $\alpha$= 0.1 (a), 
1 (b), 5 (c). The dotted lines correspond to the locus of the vertices of the frontier for 
any $T$.

\bigskip
\noindent
Fig. 5.~
Frontier in the $(x,y,T)$-space for fixed values of $x$ and 
$\alpha$= 0.1 (a), 1 (b). 
The entire plane $x+y=1$ ($\forall T$) belongs to the separable-entangled frontier.

\vfill

%\end{multicols}

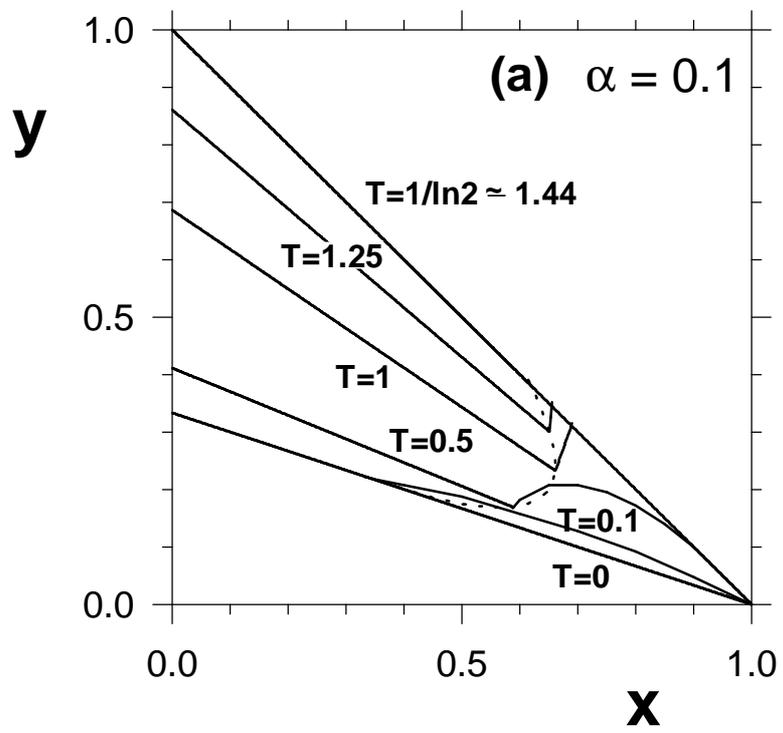
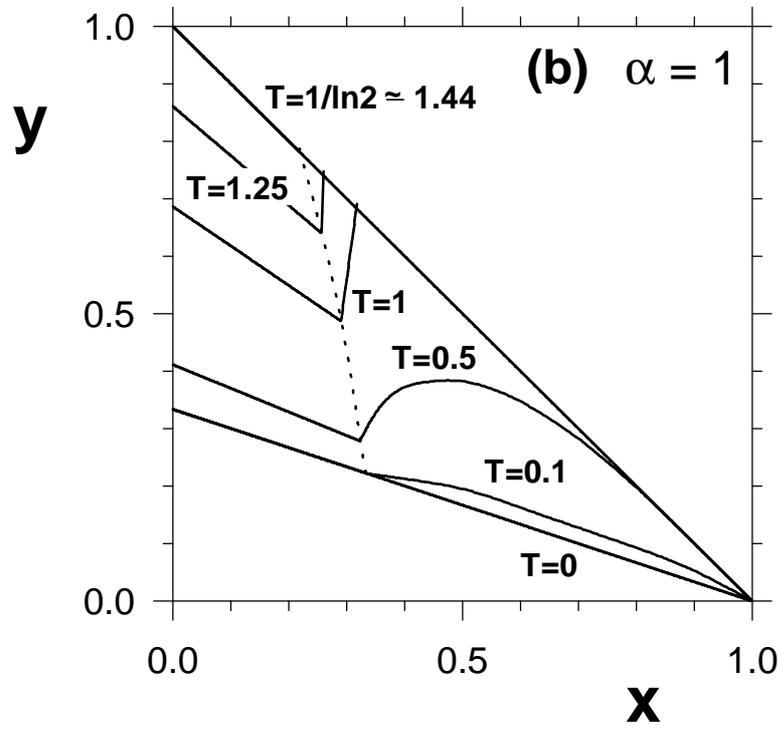
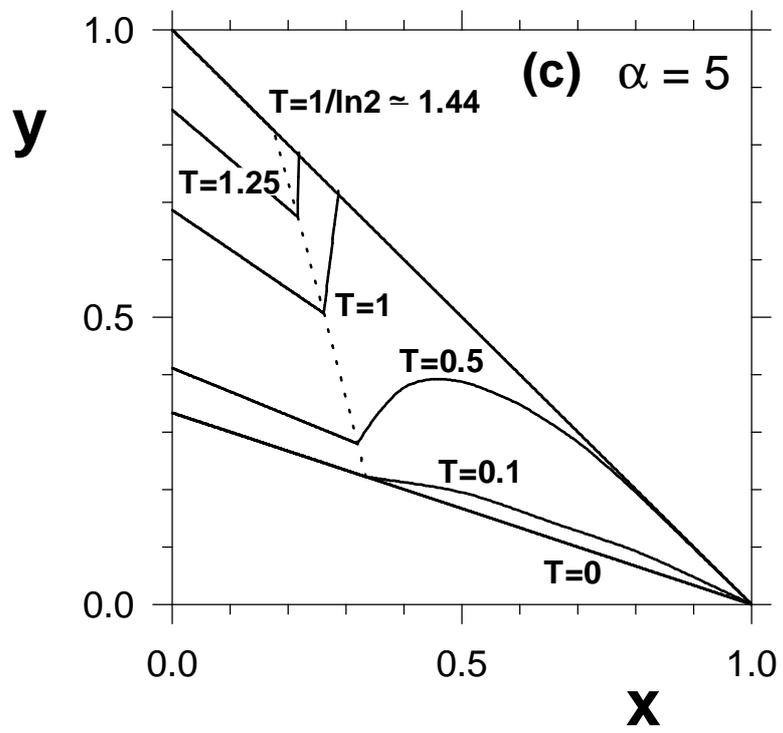
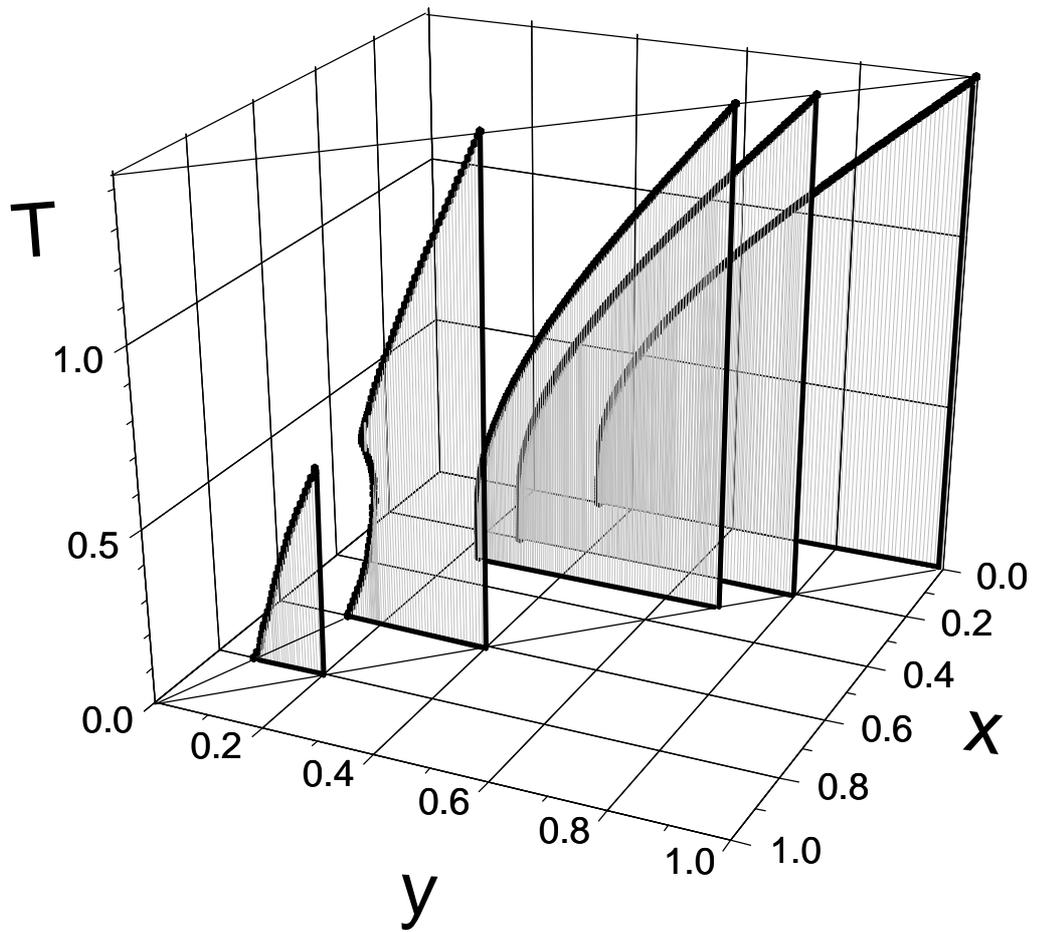
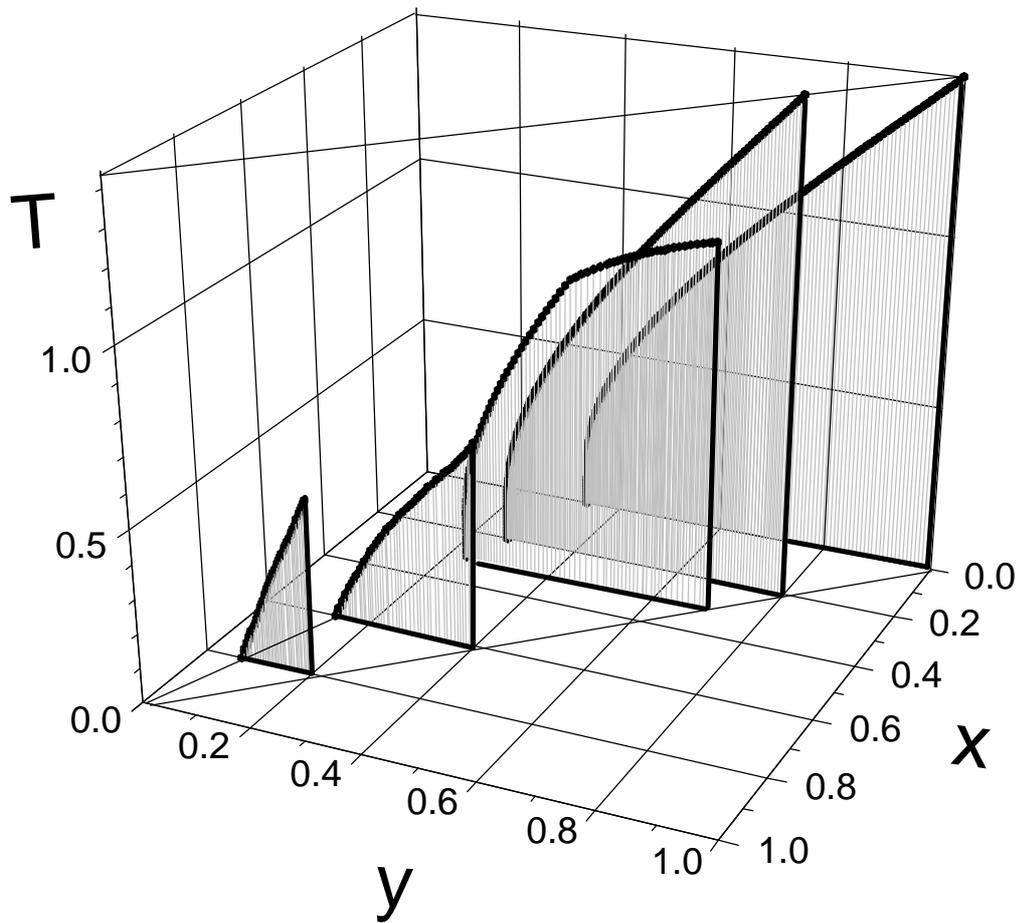
\end{document}